     \documentstyle[11pt]{article}
\expandafter\chardef\csname pre amssym.def at\endcsname=\the\catcode`\@
\catcode`\@=11
\def\hexnumber@#1{\ifcase#1 0\or 1\or 2\or 3\or 4\or 5\or 6\or 7\or 8\or
 9\or A\or B\or C\or D\or E\or F\fi}
\font\tenmsa=msam10
\font\sevenmsa=msam7
\font\fivemsa=msam5
\newfam\msafam
\textfont\msafam=\tenmsa
\scriptfont\msafam=\sevenmsa
\scriptscriptfont\msafam=\fivemsa
\edef\msafam@{\hexnumber@\msafam}
\def\emptybox{\mathrel{\mathchar"0\msafam@03}}
\catcode`\@=\csname pre amssym.def at\endcsname

\font\got=eufm10 scaled\magstep1
\font\gotscr=eufm7 scaled\magstep1
\font\gotscrscr=eufm5 scaled\magstep1
\newfam\gotfam
\textfont\gotfam=\got
\scriptfont\gotfam=\gotscr
\scriptscriptfont\gotfam=\gotscrscr
\def\got{\fam\gotfam}

\newtheorem{theorem}{\bf Theorem}
\newtheorem{lemma}{\bf Lemma}
\newtheorem{proposition}{\bf Proposition}
\newtheorem{corollary}{\bf Corollary}
\newtheorem{definition}{\bf Definition}
\newenvironment{proof}{\par\noindent{\it Proof.}}
{\kern 20pt $\emptybox$ \medskip}

\def\R{I\!\!R}
\def\Z{Z\!\!\!Z}

\title{The Orbit Method in the Finite Zone Integration Theory}
\author{P.Holod \hspace{0.3cm} and \hspace{0.3cm} S.Kondratiuk}
\date{{\normalsize Institute for Theoretical Physics,\\252143 Kiev, Ukraine\\
E-mail: mmtpitp@gluk.apc.org}}

\begin{document}
\maketitle

\begin{abstract}

A construction of integrable hamiltonian systems associated with
different graded realizations of
untwisted loop algebras is proposed.
Such systems have the
form of Euler - Arnold equations on orbits of loop algebras.
The
proof of completeness of the integrals of motion is
carried out independently of the realization of the loop algebra.
The hamiltonian systems obtained are shown to coincide with
hierarchies of higher stationary equations for some nonlinear PDE's
integrable by inverse scattering method.

We apply the general scheme for the principal and
homogeneous realizations of the loop algebra
$ sl_3(\R)\otimes{\cal P}(\lambda,\lambda^{-1}) $.
The corresponding equations on the degenerated orbit are
interpreted as the Boussinesq's and two-component modified KDV equations
respectively. The scalar Lax representation for the Boussinesq's
equation is found in terms of coordinates on the orbit applying the
Drinfeld - Sokolov reduction procedure.
\end{abstract}
\section{Introduction}

It is known that non-linear equations integrable by the
inverse scattering method admit
the
zero-curvature representation
\begin{equation}
\label{int1}
\frac{\partial U}{\partial t}-\frac{\partial V}{\partial x}+
\left[U,V\right]=0\,,
\end{equation}
where $ U(x,t) $ and $ V(x,t) $ belong to some loop algebra
$ \tilde{\got{g}} $
(algebra of Laurent polynomials with the values in a finite-dimensional
semisimple Lie algebra $\got{g} $). This representation is invariant
under the gauge transformation by the corresponding loop group.
Investigations stimulated by \cite{kacpet} resulted in a fact that the
gauge non-equivalent equations (\ref{int1}) correspond to the different
constructions of the basic representation of the loop algebra.  Each
the construction is related with a choice of the Heisenberg subalgebra,
that determines a loop algebra realization. The explicit construction
of all inequivalent graded Heisenberg subalgebras was given in
\cite{krleur}.  Any of those is determined by the finite-order
automorphism of the associated finite-dimensional Lie algebra. That
automorphisms themselves are intimately related with the conjugacy
classes of the Weyl group of the algebra. For example, in the $
sl_n(\R)\,-\, $case the Weyl group is isomorphic to the symmetric group
$ S_n $. Its irreducible representations and hence conjugacy classes
are classified by partitions of $ n $:  $$ n=n_1+n_2+\ldots+n_r\,,\quad
n_1\ge n_2\ge \ldots\ge n_r\ge 1.  $$ The partition $ n=1+1+\ldots+1 $
corresponds to the homogeneous construction, and the partition $ n=n $
is related with the principal one. When applied to the loop algebra $
\widetilde{sl_2(\R)} $, that cases lead to the hierarchies of higher
modified Korteweg - de Vries~(mKDV) equations and higher Korteweg - de
Vries~(KDV) equations respectively (cf.  \cite{hol,kis}). The
investigation of the hierarchies of equations related with the other
constructions is an actual problem (cf.  \cite{krleur,grholl,holpak}).

Let us remind that the functional phase space of a
hamiltonian system that represents a non-linear integrable PDE contains
the finite-dimensional subspaces
being invariant under the actions of hamiltonian flows generated by
all the integrals of motion of that system. That finite-dimensional
configurations arise as solutions of the higher
stationary (Novikov's) equations \cite{nov}. The
solutions provide a finite number of instability zones in the specter
of the associated linear differential operator (\,$ L\,-\, $operator).
It was shown \cite{hol} that Novikov's equations
are equivalent to the Euler - Arnold equations \cite{arn} on orbits of
the coadjoint representation of the appropriate loop group. The
time evolution is realized in that scheme naturally also. That facts
were stated considering the homogeneous realization of the
loop algebra $
\widetilde{sl_2(\R)} $. The corresponding equations were
interpreted as the higher stationary mKDV and sine- (sh-) Gordon
equations.

This article develops the scheme of
constructing the higher stationary equations
on orbits of loop algebras
of rank
$ \ge1 $ (section 2). An accent is made on
different realizations of the loop algebra. The
"intermediate" hierarchies that are related with realizations
differing from the two mentioned above are of special interest.
But in fact, the paper deals mostly with the
principal realization because of its fundamental place among the others:
every intermediately constructed affine algebra is a "modification" of
the principally realized one. Third section concerns the
examples of the principal and homogeneous realizations of the loop algebra
$ \widetilde{sl_3(\R)} $.
As a result, the equations are obtained to be
interpreted as the stationary Boussinesq's equation and the
two-component mKDV -- type equation respectively. We will also obtain
the Lax representation for the Boussinesq's equation in terms of the
coordinates on the orbit applying the Drinfeld - Sokolov reduction
procedure \cite{drsok}.

To conclude the introduction, the following should be stressed.
The orbit interpretation of the finite-zone integration theory
(presented in \cite{hol} and in this paper) allows to construct all the
theory of non-linear completely integrable PDE's in a non-traditional
way. Considering integrable hamiltonian equations on orbits
as the ground of the theory and interpreting them as
higher stationary equations for some (unknown yet) evolutionary
equations, the problem of enumeration of integrable PDE's is
reduced to an algebraic-geometrical problem to classify
the loop algebras and their orbits. It is also possible to
construct the integrals of motion for the evolutionary equations
starting from those for the stationary
equations on the orbit. The latters are got easily as expansion
coefficients (relative to the complex loop parameter) of the
Casimir functions in the enveloping algebra
of the loop algebra stated into the base of the theory.

\section{Constructing higher stationary equations:\\ the
orbit scheme}
{\bf 1. The general case.}
Let $\got{g} $ be a semisimple finite-dimensional Lie algebra of rank $ R $
and $ {\cal
P}(\lambda,\lambda^{-1}) $ the associative algebra of Laurent
polynomials with respect to the complex parameter $ \lambda $ belonging
to the unit circle. Let us consider the loop algebra
$\tilde{\got{g}}=\got{g}\otimes{\cal P}(\lambda,\lambda^{-1})$ with the
commutator:  $$ \left[\sum A_{i}\lambda^i ,\sum B_{j}\lambda^j
\right]=\sum [A_i,B_j]\lambda^{i+j}. $$
Below we will operate with the homogeneous and principal
realizations of $ \tilde{\got{g}} $. A manifestation of the difference
between them
is their different gradations with the
gradation operators $ d_h $ and $ d_p
$ respectively (see,for example, \cite{holpak}).
{}From now and later on the expressions " the realization of a loop algebra"
and "the gradation in a loop algebra" are used as equivalent.
Define the family of $Ad- \!\!$ invariant non-degenerate forms on $
\tilde{\got{g}} $:  $$
{\left\langle{A\,,B}\right\rangle}_k=\sum_{i+j=k}(A_i,B_j)\,,k\in\Z,
$$
where $ (\,,) $ denotes the Killing form in $\got{g} $.
Decompose $ \tilde{\got{g}} $ in the direct sum of two subalgebras:
$ \tilde{\got{g}}=\tilde{\got{g}}_-\oplus\tilde{\got{g}}_+ $, where $$
\tilde{\got{g}}_+=\left\{\sum_{i\ge0}A_i\lambda^i\right\}\,,\quad
\tilde{\got{g}}_-=
\left\{\sum_{i<0}A_i\lambda^i\right\}.
$$
Then $ \tilde{\got{g}}_+ $ and $ \tilde{\got{g}}_- $ is
the dual pair relative to
$ {\langle\,,\rangle}_{-1} $, with the coadjoint action
\begin{equation}
\label{star}
ad_A^*\mu=P_+[\mu,A]\,,\quad
A\in\tilde{\got{g}}_-,\,\mu\in{\tilde{\got{g}}_-}^*\simeq\tilde{\got{g}}_+,
\end{equation}
where $ P_+ $ denotes the projector onto $ \tilde{\got{g}}_+ $.
Let $ \left\{ Q_i
\right\}_1^{dim\,\gotscr{g}} $ be a basis in $\got{g} $.
Keeping in mind our further purposes, it
is more convenient to fix a dual basis
$
\left\{ Q_i^* \right\}_1^{dim\,\gotscr{g}} $ determined by  $
(Q_i^*\,,Q_j)={\delta}_{ij} $.
The finite-dimensional
subspaces
$$
M^{N+1}=\left\{\mu\in{\tilde{\got{g}}_-}^*:\mu=\sum_{l=0}^{N+1}
\sum_{i=1}^{dim\gotscr{g}}
\mu_i^lQ_i^*\lambda^l \right\} \subset{\tilde{\got{g}}_+}\,,\quad
N=0,1,2,\ldots<\infty\,, $$ where $ \mu_i^l={\left\langle\mu\,,Q_i^{-l-1}
\right\rangle}_{-1} $ are the coordinates on $ M^{N+1} $, stay
invariant under the action of $ \tilde{\got{g}}_- $. The coadjoint action
of $ \tilde{\got{g}}_+ $ is defined on $ M^{N+1} $  also.
Then $ \mu(A)={\left\langle\mu\,,A
\right\rangle}_{N+1}\,,A\in\tilde{\got{g}}_+ $, and we identify $
{\tilde{\got{g}}_+}^* $ with the subspace $ \tilde{\got{g}}_- \oplus{M^{N+1}} $
in
such a case. The coordinates on $ M^{N+1} $ can be written down as
$ \mu_i^l={\left\langle\mu\,,Q_i^{-l+N+1}
\right\rangle}_{N+1} $, and $ M^{N+1} $ stay invariant under the
action of $ \tilde{\got{g}}_+ $.

The coadjoint actions induce the family of Lie - Poisson structures on
$ M^{N+1} $:
\begin{equation}
\label{poiss}
{\{f_1,f_2\}}_{\sigma}=\sum_{l=0}^{N+1}\sum_{i=1}^{dim \got{g}}
W_{ij}^{ls}(\sigma)\frac{\partial f_1}{\partial\mu_i^l}
\frac{\partial f_2}{\partial\mu_j^s}\,,\quad\forall f_1,f_2\in{\Large\bf
C}^{\infty}(M^{N+1}),
\end{equation}
where
\begin{equation}
\label{my_first_equation}
W_{ij}^{ls}(\sigma)={\left\langle ad_{Q_{i}^{-l+\sigma}}^*\,\mu\,,
Q_j^{-s+\sigma} \right\rangle}_\sigma \,,\quad \sigma\in\Z\quad.
\end{equation}\\
\begin{definition}
Symplectic leaves of the Poisson structures
$ W(\sigma) $ will be called
generic orbits of the corresponding loop subalgebras acting on $ M^{N+1} $.
\end{definition}
The two following cases are important. Let $ {\cal
O}_-^{gen} $ denote the generic orbit of the finite-dimensional
quotient algebra $ \tilde{\got{g}}_-\left /\lambda^{-N-1}\, \tilde{\got{g}}_-
\right.
$ that acts effectively on $ M^{N+1} $. The generic orbit of
$ \tilde{\got{g}}_+\left /\lambda^{N+2}\, \tilde{\got{g}}_+ \right. $
will be denoted $ {\cal O}_+^{gen} $.

Let $ H^\nu ,\,\nu=2\,,3\,,\ldots,R+1 $, be the Casimir functions in the
enveloping algebra of
$ \got{g} $.
They are polynomials of the
variables $ \mu_k=(\mu\,,Q_k) $ on the dual $ \got{g}^* $ of $ \got{g} $.
The substitution
$ \mu_k \mapsto \mu_k(\lambda)=\sum_{l=0}^{N+1}\mu_k^l\lambda^l $
provides the continuation of $ H^\nu $ to
$ {\Large\bf C}^{\infty}(M^{N+1}) $:
\begin{equation}
\label{part}
H^\nu = \sum_{\alpha=0}^{\nu(N+1)}
h_{\alpha}^{\nu}\,\lambda^{\alpha}\,,\quad h_{\alpha}^{\nu} \in
{\bf C}^{\infty}(M^{N+1}).
\end{equation}

\begin{theorem}
\begin{enumerate}
\item
The functions $ \left\{h_\alpha^\nu\right\} $ constitute
an involutive collection in $ {\Large\bf C}^{\infty}(M^{N+1}) $,
relative to the Poisson structures $ W(-1) $ and $ W(N+1) $.  \item The
functions $ \left\{h_\alpha^\nu\right\},\,\alpha\ge(\nu-1)(N+1) $,
annihilate the Poisson structure $ W(-1) $.  \item
The functions $ \left\{h_\alpha^\nu\right\},\,\alpha=0,1,\ldots,N+1 $,
annihilate the Poisson structure $ W(N+1) $.
\end{enumerate}
\end{theorem}
\begin{proof}
Let $ {\tilde{Q}}_i^{-l+\sigma}(\sigma) $ be the tangent vector
field corresponding to the basis element
$ Q_i^{-l+\sigma} $ and the coadjoint action (\ref{star}) of
$ {\tilde{\got{g}}}_- $ (resp. $ {\tilde{\got{g}}}_+ $) for $ {\sigma}
=-1 $
(resp. $ {\sigma}=N+1 $).
Then, $ \forall f\in{\Large\bf
C}^{\infty}(M^{N+1}), $
$$
{\tilde{Q}}_i^{-l+\sigma}(\sigma)\,f(\mu)=\frac{d}{d\tau}f(Ad_{exp\,\tau
Q_i^{-l+\sigma}}^*\,\mu){\vert}_{\tau=0}=
\sum_{k,r}\frac{\partial f}{\partial \mu_{ik}^{lr}(\tau)}
\frac{d\mu_{ik}^{lr}(\tau)}{d\tau}{\vert}_{\tau=0}\,,
$$
where
$$
\mu_{ik}^{lr}(\tau)={\left\langle Ad_{exp\,\tau
Q_i^{-l+\sigma}}^*\,\mu\,,Q_k^{-r+\sigma}\right\rangle}_\sigma\,.
$$
Next,
$$
\frac{d\mu_{ik}^{lr}(\tau)}{d\tau}{\vert}_{\tau=0}={\left\langle ad_{
Q_i^{-l+\sigma}}^*\,\mu\,,Q_k^{-r+\sigma}\right\rangle}_\sigma =
                                                   W_{ik}^{lr}(\sigma).
$$
Then
\begin{equation}
\label{sstar}
{\tilde{Q}}_i^{-l+\sigma}(\sigma)=\sum_{k=1}^{dim\,g}\sum_{r=0}^{N+1}
W_{ik}^{lr}(\sigma)\frac{\partial}{\partial \mu_k^r}.
\end{equation}
Note that
$$
\frac{\partial}{\partial \mu_k^r}=
\frac{\partial \mu_k(\lambda)}{\partial \mu_k^r}\frac{\partial}{\partial
\mu_k(\lambda)} ={\lambda}^r
\frac{\partial}{\partial \mu_k(\lambda)}.
$$
By (\ref{sstar}) and
(\ref{my_first_equation}), $$ \sum_l \left(\lambda^{l+1}
{\tilde{Q}}_i^{-l-1}(-1)+ \lambda^{l-N-1}
{\tilde{Q}}_i^{-l+N+1}(N+1)\right)= \sum_{j,k}
C_{ik}^j{\mu}_j(\lambda)\frac{\partial}{\partial \mu_k(\lambda)}, $$
where $ C_{ik}^j $ denotes the structure constants of $ \got{g} $.
The $ ad^*\,- $invariance of $ H^\nu $ means
$$
\sum_{j,k} C_{ik}^j{\mu}_j(\lambda)\frac{\partial}{\partial \mu_k(\lambda)}
\,H^\nu=0,
$$
or, by previous formula,
$$
\sum_l \left(\lambda^{l+1} {\tilde{Q}}_i^{-l-1}(-1)+
\lambda^{l-N-1} {\tilde{Q}}_i^{-l+N+1}(N+1)\right)H^\nu=0.
$$
Substitute (\ref{part}) herein and
equate the coefficients at the same degrees of $ \lambda
$.  Then the following consequences arise:  \begin{equation} \label{1}
{\tilde{Q}}_i^{-l-1}(-1)\,h_{\alpha}^{\nu}=0\,,\quad \alpha\ge (\nu
-1)(N+1)\,; \end{equation} \begin{equation} \label{2}
{\tilde{Q}}_i^{-l+N+1}(N+1)\,h_{\alpha}^{\nu}=0\,,\quad 0\le\alpha\le
N+1\,; \end{equation} \begin{equation} \label{3}
{\tilde{Q}}_i^{-l-1}(-1)\,h_{\alpha}^{\nu}+{\tilde{Q}}_i^{-l+N+1}(N+1)\,
h_{\alpha}^{\nu}=0\,,\quad N+1<\alpha<(\nu-1)(N+1)\,.  \end{equation}
Second and third assertions of the theorem follow immediately from
(\ref{1}) and (\ref{2}) respectively.

The first assertion is
clear if $ \nu \ne \mu $. Let $ \nu = \mu $; (\ref{3}) leads to the
sequence of equations:
$$
{\{h_{\alpha}^{\nu},h_{\beta}^{\nu}\}}_{-1}=
{\{h_{\alpha-N-2}^{\nu},h_{\beta+N+2}^{\nu}\}}_{-1}=\cdots =
{\{h_{\alpha-m(N+2)}^{\nu},h_{\beta+m(N+2)}^{\nu}\}}_{-1},
$$
where $ m $ is a natural number. For every pair
of non-negative integer numbers $ \alpha $ and $ \beta $ that
are less then $ (\nu-1)(N+1) $ there exists
a number $ m $ such that one of the following inequalities holds:
$ \alpha-m(N+2)<0 $, or $\,\beta+m(N+2)\ge (\nu -1)(N+1) $. The first
one implies that $ h_{\alpha-m(N+2)}^{\nu}\equiv 0 $, and the second
that $ h_{\beta+m(N+2)}^{\nu} $ annihilates the Poisson structure $
W(-1) $.  In both of the cases the vanishing of the bracket $
{\{h_{\alpha}^{\nu},h_{\beta}^{\nu}\}}_{-1} $ is provided.
The involutivity of the functions $ h_{\alpha}^{\nu} $ and $
h_{\beta}^{\nu} $ with respect to $ {\{\,,\}}_{N+1}
$ can be proved in the same way.  \end{proof}\\
{\em Remark 1.\/} This proof is a realization of the "Adler scheme"
\cite{adl,adlmo}
and carried out in a gradation invariant way. A.G. Reyman and
M.A. Semenov-Tian-Shansky
proved an analogue of Theorem~1 for homogeneous gradation case applying
the $ R-$matrix technique \cite{reyman}.

By Theorem 1, the
generic orbit $ {\cal O}_-^{gen} $ is the real algebraic
manifold embedded into $ M^{N+1} $ by the constraints $ h_\alpha^\nu=
C_{\alpha}^{\nu}\,\,,\alpha\ge {(\nu-1)(N+1)} $. Next,
fixation of the functions $ h_{\alpha}^\nu,\,\alpha=0,1,\ldots,N+1 $,
determines the real algebraic manifold
being the generic orbit $ {\cal O}_+^{gen} $.
Set the hamiltonian flows of the form
\begin{equation}
\label{add}
\frac{d\mu}{d\tau_{\alpha}^\nu}=\{\mu,h_{\alpha}^\nu\}_\sigma=
ad_{dh_{\alpha}^\nu}^*\,\mu=\left[\mu,dh_{\alpha}^\nu \right]\,,
\end{equation}
on $ {\cal O}_-^{gen} $ (resp. $ {\cal O}_+^{gen} $),where
$ \alpha<(\nu-1)(N+1) $ (resp. $ \alpha>N+1 $) and
$ dh_{\alpha}^\nu $ is the differential of the hamiltonian
$ h_{\alpha}^\nu $ ($ \tau_{\alpha}^\nu $ is the corresponding
trajectory parameter).\\
{\em Remark 2.\/} Consider the Poisson structure $ W(-1) $. By
(\ref{add}),where $ \sigma=-1 $, the
coordinates $ \mu_k^{N+1} $ do not change under the action of any
hamiltonian $ h_\alpha^\nu $.
The constraints $\mu_k^{N+1}=const $ determine,
first of all, the embedding $ M^N \hookrightarrow M^{N+1} $ and,
second, the initial point of the generic
orbit $ {\cal O}_-^{gen} $. Thus, the
symplectic structure $ W(-1) $ is meant as restricted to $ M^N $.
Considering the coadjoint action of
$ \tilde{\got{g}}_-\left/\lambda^{-N-1}\, \tilde{\got{g}}_- \right. $,
the functions $ h_{\nu (N+1)}^{\nu} $ are
fixed constants on  $ M^N $ and hence must be
neglected. Similarly, $ h_0^{\nu} $
are constants on $ {M^N}'=M^{N+1}\setminus span_{\R}\{\mu_i^0\} $
(with $ \setminus $ being set minus) if
the action of $ \tilde{\got{g}}_+\left /\lambda^{N+2}\,
\tilde{\got{g}}_+ \right. $ is considered with respect to $ W(N+1) $.
\begin{lemma}
Let ${\got{g}}\simeq sl_n(\R) $.  \begin{enumerate}
\item The dimensions of the generic orbits $ {\cal O}_-^{gen} $ and $
{\cal O}_+^{gen} $ are equal to $$ dim \,{\cal O}_-^{gen}= dim \,{\cal
O}_+^{gen}=(N+1)(n-1)n. $$
\item The number of the non-annihilators on
the generic orbits is equal to
$$
\#(Ham)=\frac{(N+1)(n-1)n}{2}.
$$
\end{enumerate}
\end{lemma}
\begin{proof}
By Theorem 1 and keeping in mind {\em Remark 2\/}, the
assertions follow from the straightforward calculations.
The only point to check is the functional independence
of the $ W(-1)- $annihilators $ h_\alpha^\nu
\,,(\nu-1)(N+1)\le\alpha<\nu(N+1)\, $, in the space $ M^N $,
and the $ W(N+1)-$annihilators $ h_\alpha^\nu
\,,0<\alpha\le N+1\, $, in the space $ {M^N}' $. These facts
are proved in the next proposition.
\end{proof}
\begin{proposition}
\begin{enumerate}
\item
Let not all $ \mu_i^{N+1} $ be equal to zero.
\begin{enumerate}
\item
The
functions $ h_\alpha^\nu
\,,(\nu-1)(N+1)\le\alpha<\nu(N+1)\, $, are functionally independent
almost everywhere on $ M^N $.
\item
The functions $ h_\alpha^\nu
\,,0\le \alpha<(\nu-1)(N+1)\, $, are functionally independent
almost everywhere on $ {\cal O}_-^{gen} $.
\end{enumerate}
\item
Let not all $ \mu_i^0  $ be equal to zero.
\begin{enumerate}
\item
The
functions $ h_\alpha^\nu
\,,0< \alpha\le N+1\, $, are functionally independent on $ {M^N}' $.
\item
The functions $ h_\alpha^\nu
\,,\alpha>N+1\, $, are functionally independent
on $ {\cal O}_+^{gen} $.
\end{enumerate}
\end{enumerate}
\end{proposition}
\begin{proof}
We prove only two first statements, the
second two are proved quite similarly. If $\nu\ne\mu $, the
statements are obvious. Consider the matrix with the rows
formed by the partial derivatives of the functions $
h_\alpha^\nu\,,(\nu-1)(N+1)\le\alpha<\nu(N+1) $, with respect to the
coordinates on $ M^N $. It is quasitriangular. One can
construct the minor of order $ N+1 $ from the entries of the
matrix. That minor is not equivalent to zero in all the points of
$ M^N $ except $ \mu_i^N=0 $. So the statement 1.a is
proved. Next, consider the similar matrix of the derivatives but for $
0\le \alpha <\nu(N+1) $ and where the derivatives are meant to be with
respect to the coordinates on the orbit $ {\cal O}_-^{gen} $.  It is a
$ \nu(N+1) \times (N+1)(n-1)n $ -- matrix. There exist
not less than $ (\nu-1)(N+2) $ different minors of order $ \nu(N+1)
$ constituted by the entries of that matrix. Construct such
minors for every $
\nu $. Equating the minors obtained to zero one gets the system of the
algebraic equations determining the set of singular points $
M^N_S\subset M^N $. Its dimension is less than or equal to $ \sum_{\nu =2}^n
(\nu -1)(N+2)=(n-1)(N\frac{n+2}{2}+1) $.  So $ dim\,M^N_S< dim\,{\cal
O}_-^{gen} $, which proves the functional independence of the functions
$ h_\alpha^\nu\,, 0\le \alpha <\nu(N+1) $ almost everywhere on $ {\cal
O}_-^{gen} $. Now the statement 1.b becomes obvious.  \end{proof}
\begin{corollary}
Hamiltonian flows (\ref{add}) are integrable in the
Liouville sense.
\end{corollary}
\begin{proof}
By Liouville theorem on the complete integrability \cite{arn},
the assertion follows from Theorem~1, Lemma 1 and Proposition 1.
\end{proof}

Given
hamiltonian $ h_{\alpha}^\nu $, a Legendre-type
transformation $ { h_{\alpha}^\nu \mapsto {\cal L}( h_{\alpha}^\nu )} $
can be defined,
and the corresponding hamiltonian system
$$
\frac{d\mu}{d\tau_{\alpha}^\nu}={\left\{\mu, h_{\alpha}^\nu \right\}}_\sigma
$$
on the orbit admits the form of the Euler - Lagrange equation:
\begin{equation}
\label{second}
\frac{\delta L( h_{\alpha}^\nu )}{\delta \mu}=0 \,,
\end{equation}
where $ L( h_{\alpha}^\nu ) $ is the Lagrange function associated with
$ h_{\alpha}^\nu $.
The crucial point is that the
hamiltonian flows generated by non-annihilators $ h_{\alpha}^\nu $ are
invariantly embedded one into another. That means, for example, that
the hamiltonian flow generated by $ h_{\alpha}^\nu $ on
the orbit $ {\cal O}_-^{gen} $ can be written as
(\ref{second}), where $$ L( h_{\alpha}^\nu )=\int\!\bigl\{c_0{\cal L}(
h_0^\nu )+c_1{\cal L}( h_1^\nu )+\ldots +c_{\alpha -1}{\cal L}
(h_{\alpha -1}^\nu)+c_{\alpha}
{\cal L}(h_{\alpha}^\nu )\bigr\}d\tau_{\alpha}^\nu\qquad,
$$
with the constants
$ c_i $ being linear combinations of the annihilators determining the
orbit.
This fact enables us to identify the Euler - Arnold equation of the
form (\ref{add}) with the higher stationary equation for some
evolutionary integrable system. The hierarchy of the higher
stationary equations arise since the number $ N $ can be chosen
as large as necessary but finite. The lagrangian densities
$ {\cal L}(\,\cdot\,) $ appear as the densities of integrals of motion
for the hierarchy of evolutionary equations, and are constructed
purely algebraically without using the associated linear problem.

An additional ("time") evolution is realized as the action of any hamiltonian
flow
generated by non-annihilators  $ h_{\beta}^\mu\ne h_{\alpha}^\nu $ on
stationary trajectory points of the hamiltonian system (\ref{add}).
As the result the system of equations on the
orbit arises:
\begin{equation}
\label{new}
\frac{\partial \mu}{\partial\tau_{\alpha}^\nu}={\{\mu,
h_{\alpha}^\nu \}}_\sigma\,,\quad
\frac{\partial \mu}{\partial\tau_{\beta}^\mu}={\{\mu,
h_{\beta}^\mu \}}_\sigma\,.
\end{equation}
Its compatibility condition
has the zero-curvature representation form for the
restriction of an evolutionary equation onto the orbit:
$$
\frac{\partial h_{\alpha}^\nu}{\partial\tau_{\beta}^\mu}-
\frac{\partial h_{\beta}^\mu}{\partial\tau_{\alpha}^\nu}+
\left[dh_{\beta}^\mu,dh_{\alpha}^\nu\right]=0.
$$
{\bf 2. The $ sl_n(\R) $--case.}
Let ${\got{g}}\simeq sl_n(\R) $. Then $ R=n-1 $. Let $ E_{ij} $
denote $ n\times n $--matrix with the unit at the $ (ij)-$entry
and zeros elsewhere.
Fix the dual basis of $ sl_n(\R) $:
$$ \left\{
Q_i^* \right\}_1^{dim\,\got{g}}=\left\{H_1^*,\ldots ,H_R^*, X_1^*,\ldots
,X_{\frac{R(R+1)}{2}}^*,Y_1^*,\ldots ,Y_{\frac{R(R+1)}{2}}^*
\right\}\,,
$$
where
$$
H_i^*=E_{ii}-E_{i+1,i+1};
$$
$$
X_1^*=E_{12},X_2^*=E_{23},\ldots ,X_R^*=E_{R,R+1},
X_{R+1}^*=E_{13},X_{R+2}^*=E_{24},\ldots ,X_{2R-1}^*=E_{R-1,R+1},
$$
$$
\ldots\ldots\ldots
$$
$$
X_{\frac{R(R+1)}{2}}^*=E_{1,R+1};\quad Y_i^*={X_i^*}^T.
$$
The homogeneous and principal gradation operators are given by
$$
d_h=\lambda\frac{d}{d\lambda},
$$
$$
d_p=n\lambda\frac{d}{d\lambda}+ad_{diag\left(\frac{n-1}{2},\frac{n-1}{2}-1,
\ldots,-\frac{n-1}{2}\right)},
$$
respectively.
The
elements of the dual basis of
$ \widetilde{sl_3(\R)} $ have the
following grades with respect to $ d_h $ and $ d_p $:
$$
d_h\,\left(\lambda^k\cdot Q_i^*\right)=k\cdot \left(\lambda^k\cdot
Q_i^*\right) ,\,i=1,2,\ldots , dim\,\got{g}; $$\\ $$
d_p\,\left(\lambda^k\cdot H_i^*\right)=nk\cdot \left(\lambda^k\cdot
H_i^*\right),\, i=1,2,\ldots ,R;
$$
$$
d_p\,\left(\lambda^k\cdot X_i^*\right)=(nk+1)\cdot
\left(\lambda^k\cdot X_i^*\right),\, i=1,2,\ldots ,R; $$ $$
d_p\,\left(\lambda^k\cdot X_i^*\right)=(nk+2)\cdot
\left(\lambda^k\cdot X_i^*\right),\, i=R+1,R+2,\ldots ,2R-1\,; $$ $$
\ldots\dots\ldots\dots
$$
$$
d_p\,\left(\lambda^k\cdot X_{\frac{R(R+1)}{2}}^*\right)=(nk+R)\cdot
\left(\lambda^k\cdot X_{\frac{R(R+1)}{2}}^*\right); $$ $$
d_p\,\left(\lambda^k\cdot Y_i^*\right)=(nk-1)\cdot
\left(\lambda^k\cdot Y_i^*\right),\, i=1,2,\ldots ,R\,; $$ $$
\ldots\ldots\ldots
$$
$$
d_p\,\left(\lambda^k\cdot Y_{\frac{R(R+1)}{2}}^*\right)=(nk-R)\cdot
\left(\lambda^k\cdot Y_{\frac{R(R+1)}{2}}^*\right).
$$
The Casimir functions in the enveloping algebra of $ sl_n(\R) $ are
$ H^\nu=\frac{1}{\nu}tr\,A^{\nu}\,,
\nu=2,3,\ldots,n $, where $ A $ belongs to the dual of $ sl_n(\R) $.
\section{The higher stationary equations on orbits of the
principally and homogeneously realized loop algebra
$ \widetilde{sl_3(\R)} $}
{\bf 1. The principal realization.}
An
element $\mu\in M^{N+1} $ is given by
$$
\mu=\sum_{l=0}^{N+1}\left\{\sum_{i=1}^{2}\left(\alpha_i^{3l}H_i^*\lambda^l+
\beta_i^{3l+1}X_i^*\lambda^l+\gamma_i^{3l-1}Y_i^*\lambda^l\right)+
\beta_3^{3l+2}X_3^*\lambda^l+\gamma_3^{3l-2}Y_3^*\lambda^l\right\}\,,
$$
where the lower coordinate indices relate to the
dual basis elements of $ sl_n(\R) $, and the
upper indices denote the grades
of the corresponding dual basis elements of $ \widetilde{sl_3(\R)} $.
Consider the Poisson manifold $ (M^{N+1},W(-1)) $.

It follows from the theory of finite-dimensional Lie
algebras that an initial point of a coadjoint representation
orbit is determined uniquely by the point in the Weyl chamber in the span
of the Cartan subalgebra (cf. \cite{bourb}).
By the analogy with that, we determine an initial point of the generic orbits
fixing a point in the Weyl chamber for $ sl_3(\R) $ that is the
coefficient at $ \lambda^{N+1} $. In the case of principally
realized algebra $ \widetilde{sl_3(\R)} $, the Weyl chamber of $ sl_3(\R) $
is
(cf. \cite{kackaz}) $ span_{\R}\left\{\delta_1
{\got{W}}_1 + \delta_2 {\got{W}}_2 \right\}\,, \delta_1 ,
\delta_2 \in {\R}_+ $, where ${\got{W}}_1=E_{12}+E_{23}+E_{31},\,
{\got{W}}_2=E_{13}+E_{21}+E_{32} $.  Let us choose the initial point on
the boundary of the chamber: $ \delta_1=0\,,\delta_2=1 $. That means in
coordinate terms:  $
\beta_1^{3(N+1)+1}=\beta_2^{3(N+1)+1}=\gamma_3^{3(N+1)-2}=0\,,
\gamma_1^{3(N+1)-1}=\gamma_2^{3(N+1)-1}=\beta_3^{3(N+1)+2}=1\, $.
The embedding $ M^N\hookrightarrow M^{N+1} $ is completed
putting $ \alpha_1^{3(N+1)}=\alpha_2^{3(N+1)}=0 $.

Put $ N=0 $.
By Theorem 1, the generic orbits are fixed by
$$
\begin{array}{l}
h_1^2=\beta_1^1\gamma_1^2+\beta_2^1\gamma_2^2+\beta_3^5\gamma_3^{-2}=
C_1^2\,,\\\\
h_2^3=\beta_3^2\gamma_1^2\gamma_2^2+\beta_
3^5(\gamma_1^{-1}\gamma_2^2+\gamma_2^{-1}\gamma_1^2)=C_2^3\,,
\end{array}
$$
where $ C_1^2 $ and $ C_2^3 $ are constants.
The explicit forms of the $ W(-1)-$annihilators $ h_1^2 $ and $ h_2^3 $
provide the trivial topological structure of the real algebraic manifold
$ {\cal O}_-^{gen}\simeq {\R}^6 $.
Consider the additional degeneration of the orbit determined by
$ \gamma_3^{-2}=0 $. The functions
$$
\begin{array}{l}
h_0^2=(\alpha_1^0)^2+(\alpha_2^0)^2-\alpha_1^0\alpha_2^0+
\beta_1^1\gamma_2^{-1}+\beta_2^1\gamma_2^{-1}\,,\\\\
h_0^3=\alpha_2^0\beta_1^1\gamma_1^{-1}-\alpha_1^0\beta_2^1\gamma_2^{-1}-
\alpha_1^0(\alpha_2^0)^2+\alpha_2^0(\alpha_1^0)^2+\beta_3^2\gamma_1^{-1}
\gamma_2^{-1}\,,\\\\
h_1^3=\alpha_2^0\beta_1^1\gamma_1^{2}-\alpha_1^0\beta_2^1\gamma_2^{2}+
\beta_3^2\gamma_1^{-1}\gamma_2^2+\beta_3^2\gamma_2^{-1}\gamma_1^2\,,
\end{array}
$$
create integrable hamiltonian flows on the degenerated orbit.
Fix the $ W(-1)-$annihilators: $ C_1^2=0\,,C_2^3=1 $, and
consider the hamiltonian system
\begin{equation}
\label{ham}
\frac{d\mu}{d\tau_0^3}={\{\mu\,,h_0^3\}}_{-1}\!=
ad_{dh_0^3}^*\mu
\end{equation}
\begin{proposition}
The equation (\ref{ham})
admits the restriction onto the immovable points of the following
involution of the dual of  $ sl_3(\R) $: $ H_1^*\mapsto H_2^*\,,
H_2^*\mapsto H_1^*\,,X_1^*\mapsto -X_2^*\,,X_2^*\mapsto -X_1^*\,,
Y_1^*\mapsto -Y_2^*\,,Y_2^*\mapsto -Y_1^*\,,X_3^*\mapsto X_3^*\,,
Y_3^*\mapsto Y_3^*\,$.
\end{proposition}
\begin{proof}
The set of
immovable points is determined by the constraints
$ \alpha_1^0=-\alpha_2^0\,,\gamma_1^{-1}=
-\gamma_2^{-1}\,,\beta_1^1=-\beta_2^1. $ This and the explicit form
of (\ref{ham}) make the assertion obvious.
\end{proof}\\
Denote
$ \alpha_1^0=-\alpha_2^0\equiv\alpha\,,\gamma_1^{-1}=
-\gamma_2^{-1}\equiv\gamma\,,\beta_1^1=-\beta_2^1\equiv\beta=const $
(the latter is by (\ref{ham})).
Then (\ref{ham}) reduces to the system of two ordinary
differential equations:
\begin{equation}
\label{redham}
\left\{
\begin{array}{l}
\alpha'+\gamma+\beta\alpha=0\\\\
-\gamma'+3\alpha^2+\beta\gamma=0\,.\\
\end{array}\right.,
\end{equation}
where $ (\,\cdot\,)'\equiv \frac{d(\,\cdot\,)}{d\tau_0^3} $.
\begin{proposition}
The hamiltonian system (\ref{redham}) admits the
Euler - Lagrange form:  $$ \left\{ \begin{array}{lr} {\displaystyle
\frac{\delta L}{\delta \gamma}}=0\,&\\&\,,\qquad L=\int({\cal
L}_2+\beta {\cal L}_1)d\tau_0^3\,,\\ {\displaystyle \frac{\delta
L}{\delta \alpha}}=0\,& \end{array} \right.  $$ where the lagrangian
densities $$ {\cal L}_1=\alpha\gamma\,,\quad {\cal
L}_2=\gamma\alpha'+\alpha^3+ \frac{1}{2}\gamma^2\,.  $$
\end{proposition}
\begin{proof}
Straightforward verification.
\end{proof}\\
The lagrangian
densities $ {\cal L}_1 $ and $ {\cal L}_2 $ coincide
(up to total derivatives) with the corresponding densities of integrals
of motion
for the Boussinesq's equation (cf. \cite{mckean}). Hence, the Euler -
Arnold equation (\ref{ham}) on the degenerated orbit is interpreted as
the stationary Boussinesq's equation. Furthermore, the lagrangian
density $ {\cal L}_2 $ has the form of the Legendre - type
transformation
$$
h_0^3\mapsto {\cal L}(h_0^3)\equiv {\cal L}_2=\frac{1}{2}\left(
\gamma\frac{\partial h_0^3}{\partial\gamma}-h_0^3\right)\,.
$$
The hierarchy of higher stationary
Boussinesq's equations appears when the subspaces $ M^{N+1}\,,N=0,1,2,
\ldots <\infty $, are involved. The densities of higher integrals of
motion are calculated in the same way as for $ N=0 $.

Remind the scalar Lax representation for the Boussinesq's equation
(cf., for example, \cite{drsok}):
$$
\frac{dL}{dt}=\left[L\,,(L^{\frac{2}{3}})_+\right]\,,\quad
L=\frac{{\partial}^3}{{\partial x}^3}+u\frac{\partial}{\partial x}+v\,,
$$
with the explicit form
$$
\left\{
\begin{array}{l}
{\displaystyle \frac{\partial u}{\partial t}=-\frac{{\partial}^2 u}
{{\partial x}^2}+
2\frac{\partial v}{\partial x}
}\,,\\\\
{\displaystyle \frac{\partial v}{\partial t}=\frac{{\partial}^2 v}
{{\partial x}^2}-
\frac{2}{3}\frac{{\partial}^3 u}{{\partial
x}^3}-\frac{2}{3}u\frac{\partial u}{\partial x}
}\,.
\end{array}
\right.
$$
We are going to connect the standard unknown functions $ u $ and $ v $
with the variables on the orbit. To do that consider the coadjoint
action of the subalgebra $ \tilde{g}_+ $ on the Poisson manifold
$ \left(M^1\,,W(0)\right) $. One can easily check that the
hamiltonian flow
\begin{equation}
\label{f}
\frac{d\mu}{d\tau_1^3}={\left\{\mu\,,h_1^3\right\}}_{0}=
ad_{dh_1^3}^*\,\mu=\left[\mu\,,dh_1^3 \right]\,,
\end{equation}
coincides with (\ref{redham}) on the degenerate orbit. Since both of the $
\mu $ and the $ dh_1^3 $ belong to $ \tilde{g}_+ $ (i.e. contain only
non-negative degrees of the parameter $ \lambda $), the equation
(\ref{f}) is the stationary zero-curvature
representation in a usual form, where $ \tau^3_1=x $.
\begin{proposition}
There exists
the unique matrix
$$
L^{can}=
\left(
\begin{array}{ccc}
0&1&0\\0&0&1\\ \lambda-v&-u&0
\end{array}
\right)\,,
$$
and the lower triangular matrix $ S=S(\tau_1^3) $ with units at the
diagonal entries,
such that the following gauge transformation takes place:
$$
S\cdot \left(\frac{d}{d\tau_1^3}+L^{can}\right)\cdot S^{-1}=
\frac{d}{d\tau_1^3}+dh_1^3\,.
$$
\end{proposition}
\begin{proof}
The statement follows immediately from Theorem 3.1 in (\cite{drsok})
once the explicit form of the matrix $ dh_1^3 $ is written down.
\end{proof}\\
By this proposition, the quite simple expressions arise:
$$
u=2\alpha\,,\qquad v=\gamma\,,
$$
describing the restriction of
$ u $ and $ v $ (up to a possible rescaling) from the
infinite-dimensional phase space onto the trajectories of the equation
(\ref{f}) on the degenerate orbit.

Let us construct the equation to be interpreted as the restriction of the
evolutionary Boussinesq's equation onto the orbit. Consider the action of the
subalgebra $ \tilde{\got{g}}_+ $ on the Poisson manifold $
\left(M^2\,,W(1)\right) $.
The hamiltonian flow generated by the coefficient function $ h_2^3 $
describes
the stationary Boussinesq's equation. Set the action of the
hamiltonian flow generated by $ h_3^3 $ on the trajectories of the
hamiltonian $ h_2^3 $. Then the system of partial differential equations
$$
\left\{
\begin{array}{l}
{\displaystyle \frac{\partial \mu}{\partial \tau_2^3}=
ad_{dh_2^3}^*\,\mu}\,, \\\\
{\displaystyle \frac{\partial \mu}{\partial \tau_3^3}=ad_{dh_3^3}^*\,\mu}\,,
\end{array}
\right.
$$
holds.
Its compatibility condition has the
zero-curvature representation form for the Boussinesq's equation:
$$
\frac{\partial dh_2^3}{\partial\tau_3^3}-
\frac{\partial dh_3^3}{\partial\tau_2^3}+
\left[dh_3^3\,,dh_2^3\right]=0.
$$
{\bf 2. The homogeneous realization.} In such a case an element $
\mu\in M^{N+1} $
reads
$$
\mu=\sum_{l=0}^{N+1}\left\{\sum_{i=1}^2\alpha_i^lH_i^*\lambda^l+\sum_{i=1}^3
\left(\beta_i^lX_i^*\lambda^l+\gamma_i^lY_i^*\lambda^l\right)\right\}\,.
$$
The homogeneous realization of $ \widetilde{sl_3(\R)} $
corresponds to the Weyl chamber for $ sl_3(\R) $ defined by
(cf. \cite{frkac})
$$
span_{\R}\left\{\kappa_1\,diag\left(\frac{2}{3}\,,-\frac{1}{3}\,,
-\frac{1}{3}\right)+
\kappa_2\,diag\left(\frac{1}{3}\,,\frac{1}{3}\,,-\frac{2}{3}\right)\right\}\,,
\quad
\kappa_1,\kappa_2\in{\R}_+\,.
$$
Set the initial point on the boundary of the chamber putting
$ \kappa_1=0\,,\kappa_2=1 $ or, in coordinate terms,
$ 2\alpha_1^{N+1}-\alpha_2^{N+1}=0\,,2\alpha_2^{N+1}-\alpha_1^{N+1}=1 $.
The constraints $ \beta_1^{N+1}=\beta_2^{N+1}=\beta_3^{N+1}=
\gamma_1^{N+1}=\gamma_2^{N+1}=\gamma_3^{N+1}=0 $, determine the embedding
$ M^N\hookrightarrow M^{N+1} $.
Consider the Poisson manifold
$ \left( M^{N+1}\,,W(-1)\right) $.
The generic orbits are fixed
by Theorem 1.

Put $ N=2 $. The generic orbit $ {\cal O}_-^{gen} $ is the real
algebraic manifold determined by the following system of conics and
cubics in $ {\R}^{24} $:  $$
h_3^2=C_3^2\,,\,h_4^2=C_4^2\,,\,h_5^2=C_5^2\,,\,h_6^3=C_6^3\,,\,h_7^3=
C_7^3\,,\,h_8^3=C_8^3\,.
$$
These algebraic equations can be solved with respect to the variables
$ \alpha_1^l\,,\alpha_2^l,\,\,l=0,1,2 $. Furthermore, for every
fixed $ l $, the equations for
finding $ \alpha_1^l $ and $ \alpha_2^l $ are linear if the variables
$ \alpha_1^{l+k} $ and $ \alpha_2^{l+k}\,,\,k=1,2 $ are found before.
That implies the diffeomorphism  $ {\cal O}_-^{gen}\simeq {\R}^{18} $
, and the variables
$ \beta_i^l\,,\gamma_i^l\,,\,\,i=1,2,3,\,l=0,1,2 $, are the global coordinates
on the orbit.

Consider the integrable hamiltonian system on the orbit:
\begin{equation}
\label{fin}
\frac{d\mu}{d\tau_1^2}=\{\mu,h_1^2\}_{-1}\,.
\end{equation}
Let $ \theta:A\mapsto (-A)^T\,,\,\,\forall A\in  sl_3(\R) $, be the
Cartan involution of
$  sl_3(\R) $. Define the "extended" Cartan involution of the
loop algebra: $ \tilde{\theta}: \lambda^k A_k \mapsto
{(-1)}^{k+1}\lambda^k{A_k}^T $.
\begin{proposition}
The system (\ref{fin}) admits the restriction
onto the set of immovable points of the involution $ \tilde{\theta} $.
\end{proposition}
\begin{proof}
The assertion is proved as that of Proposition 2.
\end{proof}\\
The Euler - Arnold equation
(\ref{fin}) on such the degenerate orbit takes the form:
$$
\left\{
\begin{array}{l}
{\beta_2^2}'''+6{(\beta_2^2)}^2{\beta_2^2}'+4\beta_2^2\beta_3^2
{\beta_3^2}'+2{(\beta_3^2)}^2{\beta_2^2}'=0\,\\\\
{\beta_3^2}'''+6{(\beta_3^2)}^2{\beta_3^2}'+4\beta_3^2\beta_2^2
{\beta_2^2}'+2{(\beta_2^2)}^2{\beta_3^2}'=0\,,
\end{array}
\right.
$$
where $ (\cdot)'\equiv \frac{d}{d\tau_1^2} $. This system  can be
interpreted as a stationary two-component
mKDV -- type equation.
\\\\

\end{document}